\documentclass[pre,aps,onecolumn,showpacs,floatfix,amsmath,amssymb]{revtex4}
\usepackage{amssymb,amsfonts,amsmath}
\usepackage[dvips]{graphicx}
\usepackage{tabularx}
\usepackage{bm}        
\usepackage{mathrsfs} 
\usepackage{color}
\newcommand{\dn}{\Delta_0}
\newcommand{\dc}{\Delta_c}

\newcommand{\Mp}{M_{peak}}
\newcommand{\Hp}{H_{peak}}
\newcommand{\rd}{R_{DST}}

\newcommand{\avg}[1]{\langle #1 \rangle}

\begin{document}

\title{Shear Induced Rigidity in Athermal Materials:  A Unified Statistical Framework}
	\author{
	Sumantra Sarkar and Bulbul Chakraborty\\
	\normalsize{
	Martin Fisher School of Physics, Brandeis University,
	Waltham, MA 02454, USA\\
}}

\begin{abstract}
Recent studies of  athermal  systems such as dry grains and dense, non-Brownian suspensions have shown that  shear can lead to solidification through the process of shear jamming  in grains and discontinuous shear thickening in suspensions. The similarities  observed between these two distinct phenomena suggest that the  physical  processes leading to shear-induced rigidity in athermal materials are universal.  
We present a   non-equilibrium statistical mechanics model, which exhibits  the phenomenology of  these shear-driven transitions:   shear jamming and discontinuous shear thickening in different regions of the predicted phase diagram. Our analysis identifies the crucial physical processes underlying shear-driven rigidity transitions, and clarifies the distinct roles played by shearing forces and the density of grains.  
 \end{abstract}
\maketitle

\section{Introduction}
Athermal materials such as dry grains and dense non-Brownian suspensions can respond to shear by organizing into  structures that support the imposed load~\cite{Cates1998}:    a process that has been termed shear-jamming (SJ) in grains~\cite{Bi_nature,granmat_jie,ren_dijksman,Sarkar2013}, and discontinuous shear thickening (DST) in suspensions~\cite{Brown2013,Morris,Heussinger2013,Wyart_Cates,Fernandez2013,Fall:2008ys,Fall:2009zr,Fall:2010vn}.   
The nature of this self-organization process has  been intensely investigated in recent months, and striking similarities have been observed between the two  transitions. This is remarkable since the SJ transition occurs through a quasistatic process  and refers to static states of particles interacting via purely repulsive contact interactions~\cite{Bi_nature,ren_dijksman,Sarkar2013}, and  DST occurs through a dynamical process that creates non-equilibrium steady states (NESS) of particles interacting via hydrodynamic and contact interactions~\cite{Morris,Mari_long}.    The single most important trigger for these transitions has been identified as the proliferation of frictional contacts~\cite{Bi_nature,JieThesis2013,Morris,Mari_long}. 
For a range of packing fractions, $\phi_s < \phi < \phi_J$, below the isotropic jamming density, $\phi_J$,  quasi static shearing causes frictional grains to come into contact leading to the SJ transition~\cite{Bi_nature,ren_dijksman,Sarkar2013}.   In a similar range of $\phi$,  athermal suspensions exhibit DST as increasing shearing rate leads to a loss of lubrication forces and increasing number of  frictional contacts~\cite{Morris,Wyart_Cates}.   
\begin{table*}[htbp]   
\resizebox{\textwidth}{!}{
\includegraphics{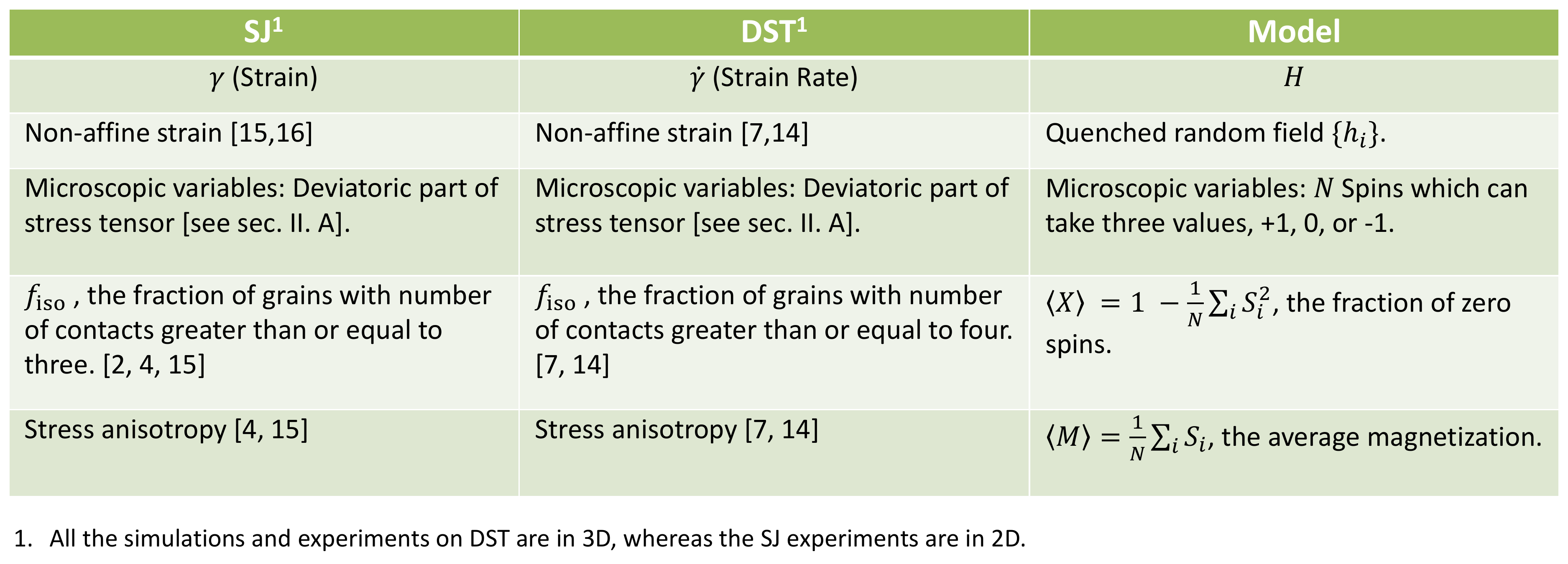}
 }
 \label{table:model}
 \caption{This table demonstrates the mapping between model parameters (Model) to the physical parameters  controlled or measured in  shear jamming (SJ)  experiments and discontinuous shear thickening (DST) simulations.}
 
 \label{summary}
\end{table*}

Lattice models have a venerable history of identifying the core physical mechanisms driving phase transitions, and finding commonalities between seemingly disparate systems.  We have constructed a  {\it non-equilibrium}, driven, disordered  model that  focuses on the process of formation and rearrangement of frictional contacts under driving by a field.   
In contrast to studies that interrogate the microscopic mechanisms leading to the SJ and DST transitions~\cite{Wyart_Cates,Sarkar2013},  we analyze an effective theory that  is built on the premise that the driving field, either strain ($\gamma$) or strain rate ($\dot \gamma$), increases shear stress {\it and} promotes the formation of frictional contacts.     We examine the consequences of the interplay between the driving field and the underlying disorder of the contact network on the development of  a robust, force-bearing network.   The model focuses solely on  the force network: changes in the network of frictional contacts with their associated tangential forces strongly affect the viscosity of suspensions in the DST regime~\cite{Morris,Mari_long,Wyart_Cates}.   
\begin{figure}[htbp]
\centering
\includegraphics[width=0.7\textwidth]{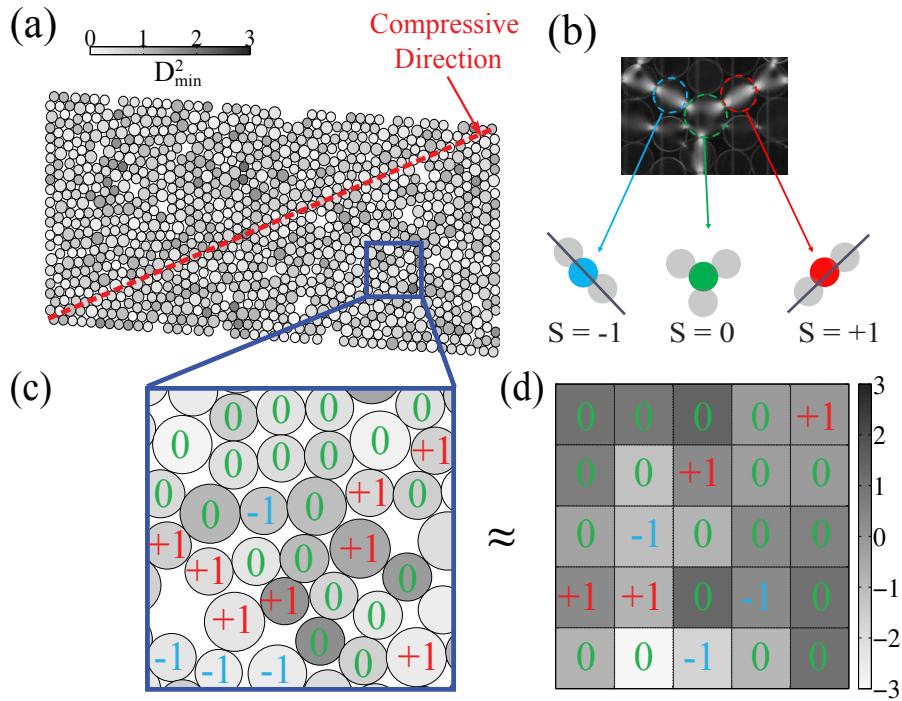}
\caption{\textbf{Mapping to spin model}  (Color online) (a) A typical sheared packing undergoing the SJ transition~\cite{JieThesis2013}, color coded according to the strength of the non-affine strain($ D^2_{min} $~\cite{Utter2008}) at a grain. (b) Mapping to spin 1 Ising variables: grains with more than $3$ contacts (green) are assumed to have $ S = 0 $ (stress anisotropy below threshold) , and grains with $2$ contacts have either $ S = 1 $ (red) or $S=-1$ (blue), depending on whether the contact is aligned along the compressive (yellow broken line in (a)) or dilational direction. (c) Enlargement of a small section of (a) illustrating the grain-spin mapping.  (d) A schematic configuration of the spin model on a square lattice, color coded  by the strength of $h_i$, which represents the non-affine strain at site $i$.  The external field $H$ is not shown. }
 \label{mapping}
 \end{figure}
 The mapping between the parameters defining the model and and the physical parameters defining and controlling force networks in the SJ and DST transitions are summarized in Tabel ~\ref{summary}.   In the next section we develop the model starting from a rigorous mapping of grain-level stresses to spins.

\section{Model}  
\subsection{Rigorous Mapping}
The tensor representing the stress state of a grain can be divided into an completely isotropic part that defines hydrostatic pressure and a deviatoric part that represents normal and shear stresses.   The deviatoric part can be represented as an element of a vector space~\cite{Ren_Behringer_APS}.  Illustrating in 2D, the stress tensor of a grain, which is symmetric since the grain is torque balanced, can be written as:
\begin{eqnarray}
\hat \sigma &=& \begin{pmatrix}\sigma_{xx} & \sigma_{xy} \\ \sigma_{xy} & \sigma_{yy} \end{pmatrix}\\ 
&=&  P\begin{pmatrix}1 & 0 \\ 0 & 1 \end{pmatrix}  + \Sigma_N \begin{pmatrix}1 & 0 \\ 0 & -1 \end{pmatrix} + \tau \begin{pmatrix} 0 & 1\\ 1 & 0 \end{pmatrix} ~,
\label{stress}\\
&=& P \hat{\sigma}_1 + \Sigma_N \hat{\sigma}_2 + \tau\hat{\sigma}_3\\
\end{eqnarray}
where $P = \frac{(\sigma_{xx }+ \sigma_{yy})}{2}$ is the hydrostatic pressure, $ \Sigma_N = \frac{(\sigma_{xx }- \sigma_{yy})}{2}$ is the normal stress, and $ \tau = \sigma_{xy}$ is the shear stress.   The deviatoric part of the stress, which excludes this hydrostatic part, is therefore an element of a  2D vector space spanned by two $2 \times 2$ matrices, $ \hat{\sigma_2} $ and $ \hat{\sigma_3} $~\cite{Ren_Behringer_APS}.   The  components of the vector are the normal stress, $ \Sigma_N$, and the shear stress, $\tau$, and the length, $ \sqrt{\Sigma_N^2+\tau^2} $,  provides a measure of the stress anisotropy of each grain.       

The stress state of a grain is influenced by the local strain arising from the displacement of the neighboring grains. The displacement of the grains comprises of a homogeneous part, which can be characterized by a set of affine transformations and an inhomogeneous part, called \textit{non-affine} displacement, which cannot be described through a series of affine transformation of the grain coordinates. The non-affine displacements are best characterized by a measure called $ D^2_{min} $, first introduced by Falk and Langer ~\cite{Falk1998}.  To calculate $ D^2_{min}$,  one measures  the actual displacements of the grains, and then chooses an optimum affine strain tensor $ \epsilon_{ij}  $,  which minimizes the mean squared deviation of the actual displacement from a homogeneous displacement due to the strain tensor.   This minimized deviation is referred to  $ D^2_{min} $. The   $ D^2_{min} $ measure has been applied to characterize the non-affine displacements in granular experiments~\cite{Utter2008}, and  shows that the non-affine strains follow a Gaussian distribution with mean approximately zero. Additionally, it has been found~\cite{JieThesis2013} that the deviatoric stress vectors interact with these non-affine strains in a manner similar to how magnets interact with spins. Thus, the continuous vectors, $ (\Sigma_N,\tau) $, can be imagined as continuous spins. The vector sum of these spins maps to the global deviatoric stress tensor, and the magnetization measures the stress anisotropy of the global stress tensor. 

\subsection{Mapping to a Spin-1 Ising Model}
Though rigorous, analyzing the properties of such a continuous spin model with variable lengths, in the presence of a random field is difficult.  We, therefore,  use a threshold to map the grain-level stress to a spin 1 Ising model. Let us define $ \Sigma_{dev} = \sqrt{\Sigma_N^2+\tau^2} $. If, $ \Sigma_{dev}/P << 1 $, we map the grain to $ S = 0 $, otherwise we map it to $ S = \pm 1 $, depending on whether the grain points along or perpendicular to the compressive strain direction. The mapping is illustrated in Fig. ~\ref{mapping} for  the two-dimensional (2D)  SJ system. In such a system, $S_i = \pm 1$ represent grains with two contacts, which have strong stress anisotropy( $ \Sigma_{dev}/P \sim 1 $) and $S_i=0$ represent grains with more than two contacts, which have a nearly isotropic stress tensor( $ \Sigma_{dev}/P << 1 $) and connect chain-like force networks.  A similar mapping applies to the three-dimensional (3D) DST systems with $S_i=0$ referring to grains connecting chain-like networks~\cite{Morris}.


We envision the SJ and DST processes as ones where the contact network and grain-level stresses reach a force and torque balanced state in the presence of driving~\cite{Sarkar2013,Morris,Mari_long}.  We model this by the zero-temperature, single-spin flips, energy minimizing dynamics of the  energy function (Fig. ~\ref{mapping}) ~\cite{Vives1994,Vives1995,Sethna2004}:
\begin{eqnarray}
  \mathcal{H}\!\!    = \!\! -J \!\!\sum_{<i,j>} S_{i} S_{j} -\!\! \sum_{i} h_{i } S_{i} -\!\! H \sum_{i} S_{i}  
  +  \Delta (H) \sum_{i} S_{i^{}}^{2} ~
 \label{eq:model}
\end{eqnarray}

In SJ experiments, it is known that as the imposed strain, $ \gamma $, is increased, the fraction of grains with small stress anisotropy ($ S = 0 $ in our model) increases. In DST, it is the strain rate that plays the same role.   Our viewpoint is that this is the primary effect of $ \gamma $ or $ \dot{\gamma} $. So we map $ H $ to $\gamma $ for SJ and to $ \dot{\gamma} $ for DST, whereas spin flips map to rearrangements of the contact network of particles. The $ H $ dependent chemical potential $ \Delta(H) $ incorporates the effect of $\gamma$ ($\dot \gamma$)  on the fraction of  grains with small stress anisotropy in SJ (DST).  In these systems with shear-induced rigidity, this fraction increases with increasing driving, and we therefore restrict our analysis to  increasing functions of $H$.  However, the model is more general and can address other scenarios. 
  
As seen in Fig. ~\ref{mapping},  shearing leads to significant non-affine displacements:  displacements that are inhomogeneous, and cannot be described by any type of homogeneous deformation of the unstrained state ~\cite{Utter2008}.       
These are represented by the random magnetic field  $h_i$, at every site.    SJ experiments indicate that the distribution of the $ D^2_{min} $ depends  on $\phi$ but evolves little during the shear-jamming process~\cite{Sarkar2013,JieThesis2013}, therefore, we treat the $h_i$ as a   {\it quenched } random field chosen from a Gaussian distribution with zero mean and standard deviation $ R $. In the granular systems, the constraints of mechanical equilibrium  introduce effective interactions between the stress tensors of grains. In a force chain where every grain has only two contacts, the anisotropy of the stresses of grains in the chain are highly correlated~\cite{Sarkar2013}. We model this effective interaction by a ferromagnetic interaction between spins. 

We are interested in understanding the effects of shear in creating robust force networks through the introduction of frictional contacts.   Our model, therefore,  differs from other driven-disordered models in the class of Eq. ~\ref{eq:model}~\cite{Sethna2004} in one crucial respect:  the external field controls the average population of  $S_i=0$ sites through  a chemical potential ($\Delta (H)$).  
For the current study, we model the $H$ dependence of $\Delta$ as $ \Delta = \alpha|H| + \Delta_0$, which is the simplest that admits an increase in the concentration of $S_i =0$   with the magnitude of the driving field.   
In this work, we focus on the aspects of the model that are relevant to shear-induced rigidity, however,  the statistics of avalanches and the yielding behavior exhibit interesting new features, which will be studied in the future.


For $N$ spins, we define two global order parameters:
\begin{equation}
\langle X  \rangle=  1 - \frac{1}{N} \sum_{i} \langle S_{i}^{2} \rangle ~ \rm{and}  ~  \langle M \rangle  =  \frac{1}{N} \sum_{i} \langle S_{i} \rangle
\label{eq:order}
\end{equation}
 Here, $\langle X \rangle $ corresponds to the fraction of grains with isotropic stress tensors ($f_{iso}$), and $\avg M$ corresponds to the  stress anisotropy (contact-stress anisotropy) in  SJ  (DST).
The zero-temperature dynamics samples the metastable states of this disordered model, which we associate with the force networks sampled in the SJ and DST processes.   
In the SJ  context, the $H$ history represents a $\gamma$ history, while in the DST context, it  represents a $\dot \gamma$ history.  Since there are no thermal fluctuations in our model,  averages ($\avg {\cdot}$ ) are over metastable states corresponding to different realizations of the quenched disorder field, $\lbrace h_i \rbrace$. To simplify notation, we eliminate the $\avg {\cdot}$ symbol in the following.

Starting from a metastable state, if $H$ is changed adiabatically such that all spins that can lower their energy by flipping do so,  there could  be a range of $H$ over which the original state is stable.   At a certain $H$, however, a threshold is crossed at some site $i$ (determined by the $h_i$ and the effective field $\sum_j J_{ij} S_j)$ and that spin changes its state.  This, in turn, could lead to the threshold being crossed at other sites, creating a cascade of spin flips in an avalanche~\cite{Sethna2004} until a new metastable state is reached.   

In the granular context,  we envision this exploration of  metastable states to correspond to exploration of force networks  that are in local mechanical equilibrium under driving.  
The SJ process is quasistatic and since the non-affine strain field is observed to evolve only weakly over the range of $\gamma$ probed by the experiments, there is a clear correspondence between  the sampling of metastable states in the model and the force networks in the granular assembly.  In DST, 
however, one studies time averages in the NESS at a given $\dot \gamma$.   The correspondence between the ensemble average over $\lbrace h_i \rbrace$ and the time average is valid if the NESS dynamically samples non-affine strains with Gaussian statistics, and  if the time to reach a force and torque balanced state is much shorter than the relaxation time of the non-affine strain field.  These assumptions are validated in simulations~\cite{Mari_private}.   The adiabatic assumption implies that $\dot \gamma$ is ramped up slowly compared to  microscopic time scales~\cite{Brown2013}.       
 {\it A priori}, it is not clear what experimental knob can be turned to tune $R$.  However,  there are strong arguments presented below, based on  comparing predictions of our model to existing experimental and numerical observations, linking increasing  $\phi$ to a reduction in $R$.  A scaling description of DST has been constructed by invoking  a stress-scale dependence of  the packing fraction at which the viscosity diverges~\cite{Wyart_Cates}.  We relate the stress-scale to $X(R,H)$, and therefore, in our approach it is the stress scale that is controlled by $\phi$, through $R$,  and  by $\dot \gamma$.  If the dominant effect of $\phi$ on the force network is through the statistics of the non-affine strain field, then the two approaches should yield similar results.   Below, we will establish specific $\phi \rightarrow R$ mappings in the SJ and DST regimes by comparing our predictions to experiments, simulations, and the scaling theory.
 

\section{Results}
\subsection{Meanfield Solution of the Model}
To solve the spin-1 Ising model under meanfield(MF) approximation, we observe that the order parameters can be represented through the probability of finding a particular value of spin at a particular lattice point:  
\begin{eqnarray}
 1 -  X & = & \frac{1}{N} \sum_{i} \avg {S_{i}^{2}}= P(S_i = 1)+ P(S_i = -1)  \nonumber \\ 
  M & = & \frac{1}{N} \sum_{i}\avg{ S_{i} }= P(S_i = 1) - P(S_i = -1) 
  \label{order_parameter}
\end{eqnarray}
where $ P(S_i = x) $ measures the probability that the $ i^{th} $ spin takes the value $ x$ ($\pm 1$ or 0). Also, in the MF approximation, the energy of a spin $ S_i $ is given by:
\begin{equation}
E(S_i) = -JMS_i - HS_i -h_iS_i +\Delta S_i^2
\end{equation}  
Therefore, 
\begin{eqnarray}
E(S_i = 1)\equiv E_1 &=& _-(JM+H+h_i) + \Delta  \nonumber \\
E(S_i = -1)\equiv E_{-1} &=& (JM+H+h_i) + \Delta \nonumber \\
E(S_i = 0) &=& 0
\end{eqnarray}
In our zero-temperature dynamics, a spin $ S_i $ will be in the +1 state if $ E_1 < 0$, and $E_1 < E_{-1} $. This condition is satisfied if 
\[ h_i > \begin{cases}
 \Delta-JM-H & \text{if } \Delta > 0 \\
 -JM - H & \text{if } \Delta \le 0
\end {cases} 
\]
whence
\begin{equation}\label{eq:P1}
P(S_i = 1)\equiv P(1) = \begin{cases}
\frac 1 2 erfc\left(\frac{\Delta-H-JM}{\sqrt 2 R}\right) & \text{if } \Delta > 0\\
\frac 1 2 erfc\left(\frac{-H-JM}{\sqrt 2 R}\right) & \text{if } \Delta \le 0\\
\end{cases}
\end{equation}
A similar calculation yields: 
\begin{equation}\label{eq:Pn1}
P(S_i = -1)\equiv P(-1) = \begin{cases}
\frac 1 2 erfc\left(\frac{\Delta+H+JM}{\sqrt 2 R}\right) & \text{if } \Delta > 0\\
\frac 1 2 erfc\left(\frac{H+JM}{\sqrt 2 R}\right) & \text{if } \Delta \le 0\\
\end{cases}
\end{equation}
Using these probabilities, and the definition of $ M$  and $X $ (Eq. ~\ref{order_parameter}), we obtain: 
\begin{eqnarray}
M &=  &\begin{cases}
    \frac 1 2 \left[ erf\left( \frac{\Delta (H) +H+M}{\sqrt 2 R}\right)\right.\\\left.\quad-  erf\left(\frac{\Delta (H) -H -M}{\sqrt 2 R}\right) \right] &  \Delta > 0 \\[6pt]
    erf\left(\frac{H+M}{\sqrt 2 R}\right)     &  \Delta \leq 0
    \label{eq:M}
  \end{cases}\\
X & =  & \begin{cases}
   \frac 1 2 \left[ erf\left(\frac{\Delta (H) +H+M}{\sqrt 2 R}\right)\right.\\\left.\quad+  erf\left(\frac{\Delta (H) -H -M}{\sqrt 2 R}\right) \right] &  \Delta > 0 \\
   0 &  \Delta \leq 0
  \end{cases}
   \label{eq:X}
\end{eqnarray} 
Here {\it{erf}}  and {\it{erfc}} are the error function and the complementary error function, respectively. Special cases and several important aspects of the MF solution are detailed in the appendix.
 
\subsection{Meanfield Phase Diagram}    
\begin{figure}[hbtp]
    \centering
 \includegraphics[width=0.5\textwidth]{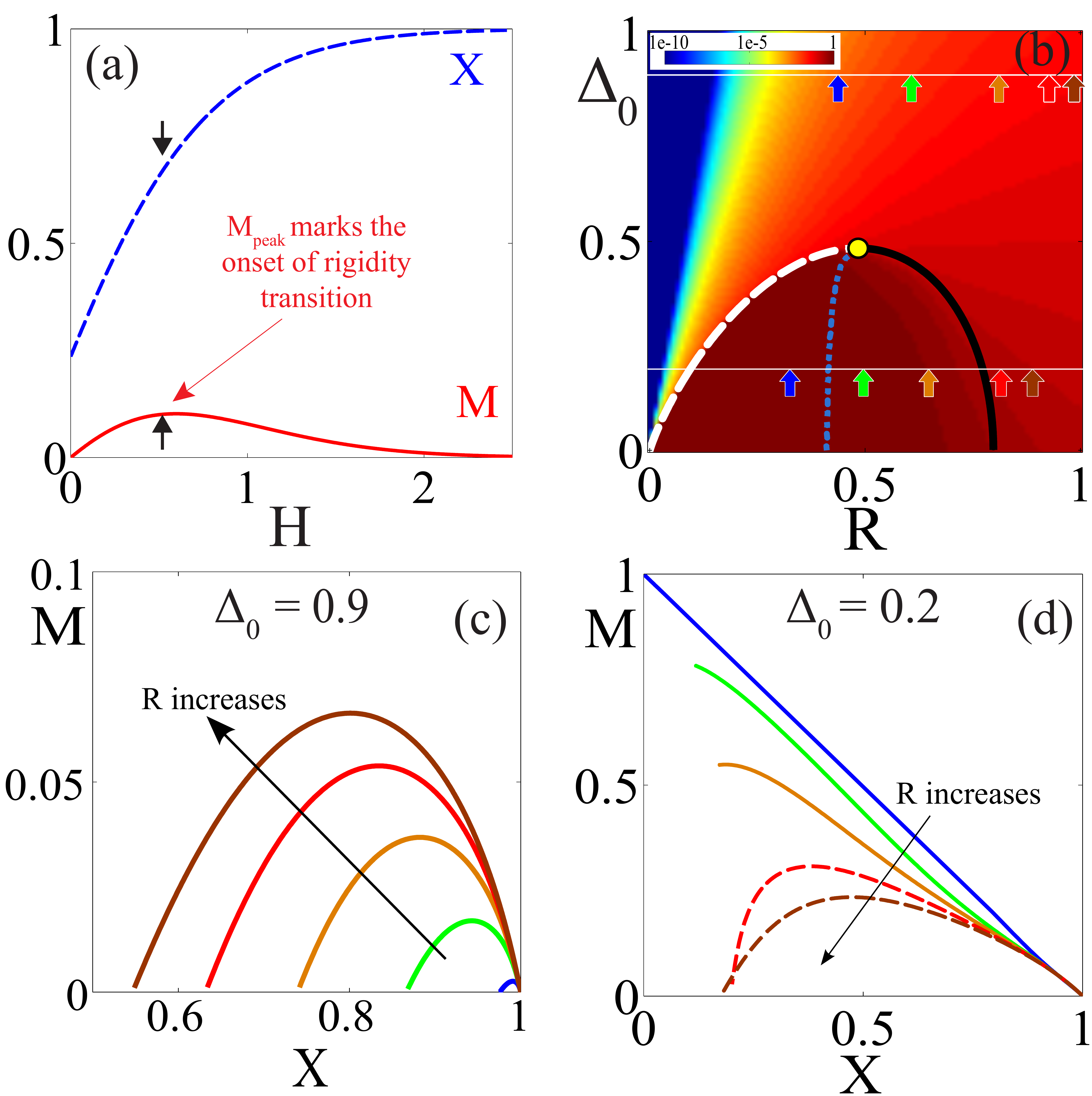} 
  \caption{ (Color Online) (a) A typical forward shear trajectory from the model for $\alpha=4$ demonstrating the appearance of $M_{peak}$, which is  concomitant with saturation of $  X $. (b) Meanfield phase diagram for $\alpha=4$:    The colorbar indicates $ \Mp $.   The critical point $(\dc, R_c)$ (yellow circle) marks the end point of  three transition lines (see text): $R_t (\dn)$ (black), $\rd (\dn)$  (light blue dotted ) and $R_m (\dn)$ (white dashed).  Detailed description of $ R_t $, $ R_m $, and $ R_{DST} $ is presented in the appendix. (c) \& (d) $  M ( X ) $  at different values of $R$  and $\dn$ can be used to characterize and distinguish between different shear induced rigidity transitions.  The different colors correspond to the values of $R$ indicated in the phase diagram.    For $ \Delta_0 > \Delta_c $ (c), $ M$ vs $ X$  is  non-montonic  whereas for $ \Delta_0 < \Delta_c $ (d), the functional form changes from non-monotonic to monotonic as $ R $ is decreased.   }
 \label{model}
 \end{figure}

Meanfield calculations of $X$ and $M$ along  a forward trajectory, with monotonically increasing $H$  (Fig. ~\ref{model}(a)) suffice to illustrate that the phenomenology of both the SJ and DST transitions are realized in the model. The meanfield phase diagram in the  $R-\Delta_0$ space is shown in Fig. ~\ref{model}(b). Increasing $\Delta_0$ corresponds to higher average concentration of zero spins at $H=0$,  corresponding to larger values of $f_{iso}$ at zero driving field.   In the SJ system~\cite{Sarkar2013},  $f_{iso}(\gamma =0) \simeq 0.2$, which maps on to the upper part of the phase diagram in Fig. ~\ref{model}(b).  In DST, however, there are a vanishing number of frictional contacts at $\dot \gamma =0$~\cite{Morris,Mari_long}, which  maps these systems to the lower part of the phase diagram.  The $R$ to $\phi$ mapping discussed earlier implies that $\phi$ decreases from left to right in Fig. ~\ref{model}(b). 

Another parameter that influences the model phase diagram is $\alpha$, the rate of increase of $\Delta$ with $H$.  As shown in Fig.~\ref{alpha} in the appendix, only the $\alpha >1 $ protocols lead to a monotonic increase of  $X(H,R)$ for all $R$, a feature of the number of frictional contacts in both SJ and DST.    We, therefore  restrict our analysis to $\alpha > 1$, and unless otherwise stated the results presented are for $\alpha=4$.   For quantitative comparisons to experiments and simulations, one should obtain  $\alpha$ by comparing the meanfield predictions for $X(H,R) $ to the increase of $f_{iso}(\gamma)$ ($(\dot \gamma)$) in SJ (DST) systems at different $\phi$.   



The qualitative differences between different regions of the phase diagram are  best characterized by  $ M( X)$, which maps on to the dependence  of stress anisotropy on $f_{iso}$.   
As shown in Fig. ~\ref{model} (c),  for $\dn >> \dc$,  $ M( X)$ has a peak ($M_{peak}$) at  $X_{peak} (\dn,R)$, which approaches $1$ as $R$ is decreased, while at the same time $M_{peak} (\dn,R) \rightarrow 0$.  This  prediction of the model is borne out by experimental SJ results, which show the same behavior with increasing $\phi$.   The weak dependence  of $ M( X,R)$ on $R$ for  $X >> X_{peak}$ is consistent with experiments~\cite{JieThesis2013}, where this regime has very weak dependence on $\phi$.  
In the limit of small $\dn$, the DST regime of the model, the functional form of  $ M( X)$  changes  with $R$, as shown in Fig. ~\ref{model} (d)).  As we discuss below, DST occurs in the $\phi$ range corresponding to $R \simeq R_{DST}$, where $M$ is a monotonically decreasing function of $X$, which explains the monotonic decrease of the  stress anisotropy with $f_{iso}$ observed in numerical  
simulations~\cite{Morris,Mari_long}.    

 \begin{table*}[htbp]   
 \resizebox{0.8\textwidth}{!}{
 \includegraphics{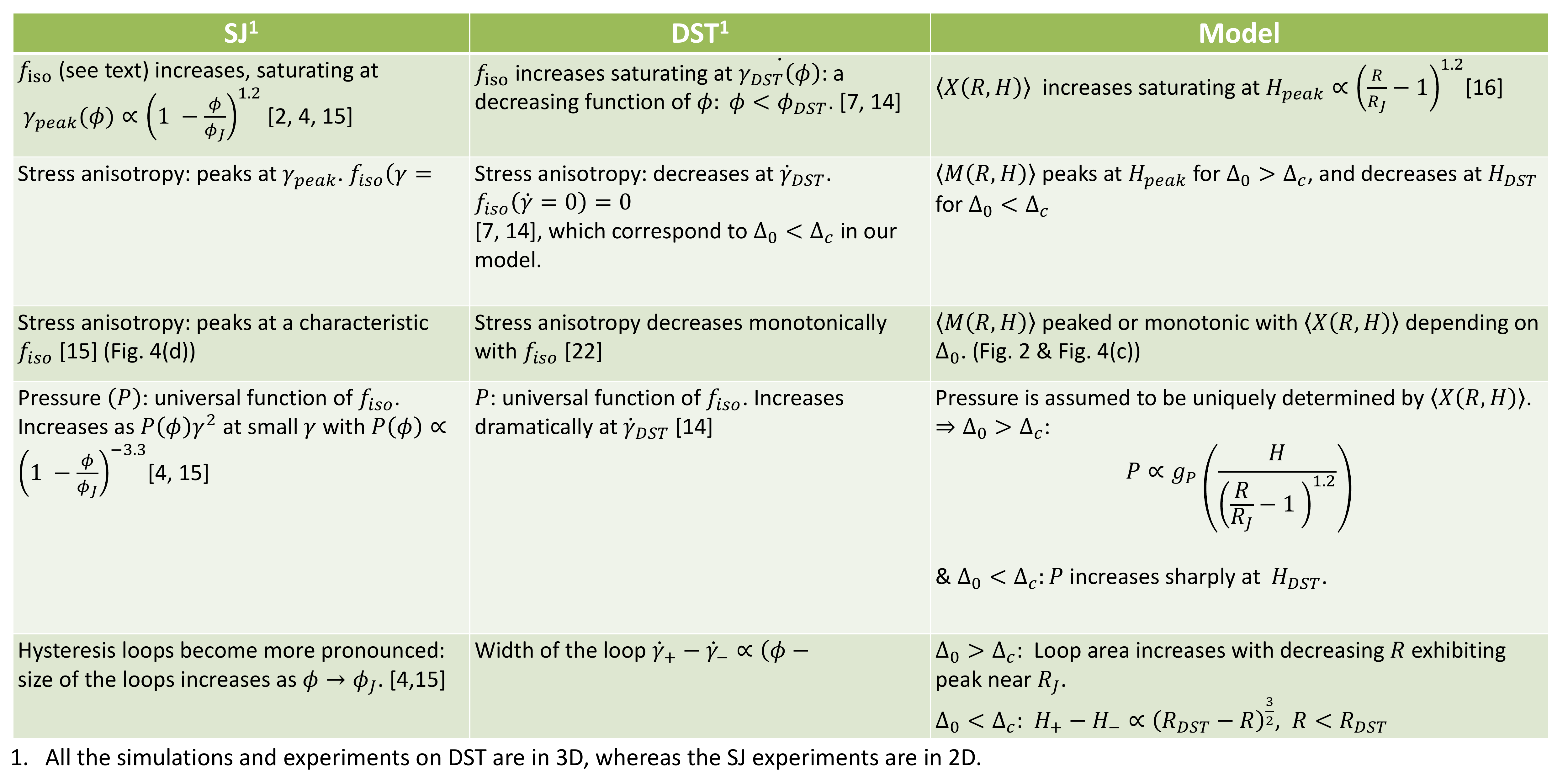}
 }
 \caption{A summary of the key predictions of the model compared to the observed phenomenology of SJ and DST transitions.}
 \label{table:results}
 \end{table*}

\subsection{Scaling and Hysteresis in SJ Regime} 
\begin{figure}[hbtp]
      \centering
   \includegraphics[width=0.5\textwidth]{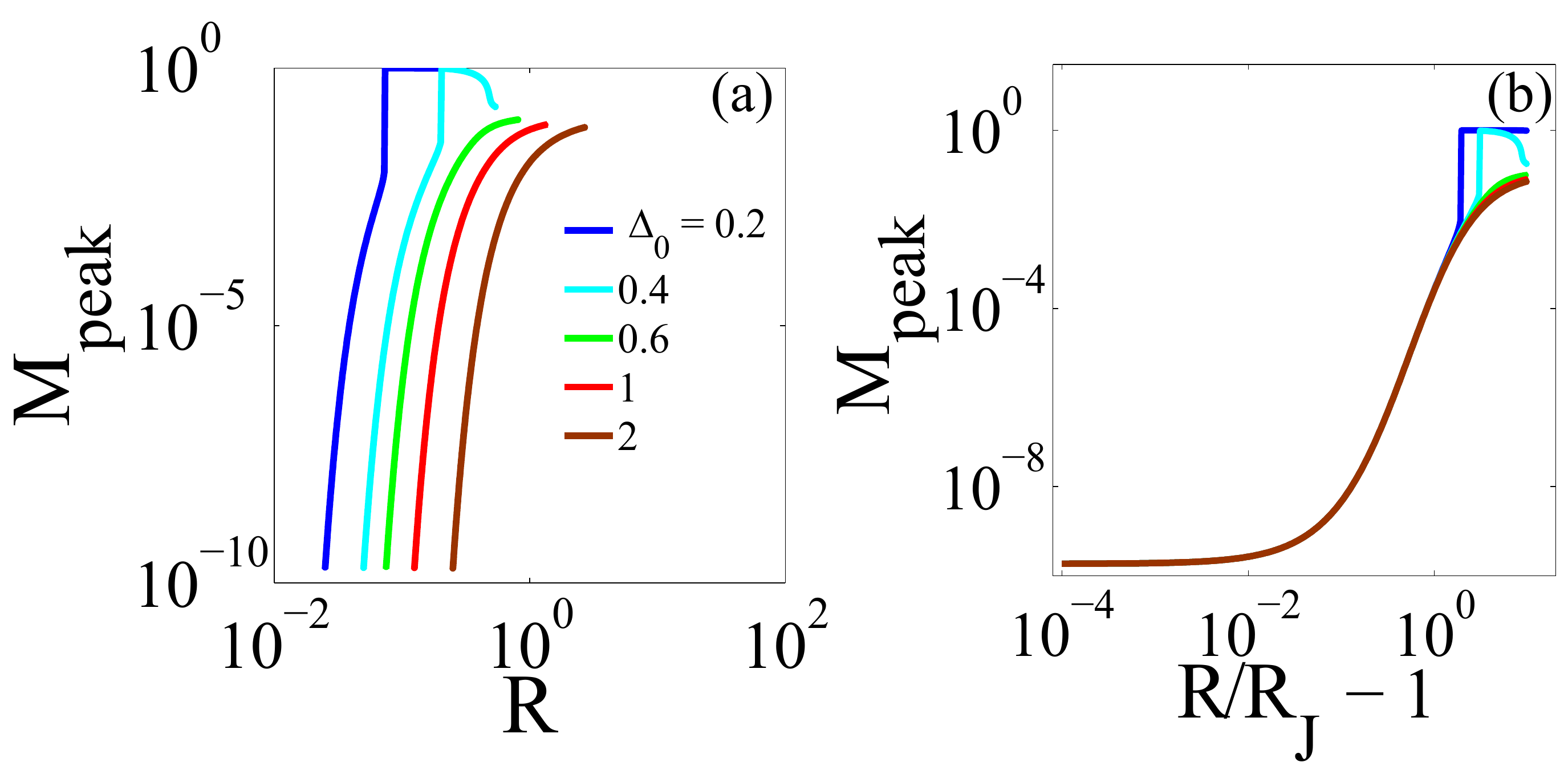} 
   \caption{(Color Online)  \textbf{Scaling of $ M_{peak} $:}  The system achieves peak anisotropy at the rigidity transition. (a) For $\dn > \dc$,  the peak value $ M_{peak} $  is continuous but for $\dn < \dc$, $M_{peak}$ has a discontinuity at $R_m$ as discussed earlier.  (b) $M_{peak}$ has a scaling form as a function of $R/R_J$  with $ R_J(\dn) = \dn/6$. For $\dn < \dc$ , the scaling form is valid for   $ R \le R_m$, and the discontinuity at $R_m$  is evident in this scaling plot. The peak anisotropy at $ R_J$ is $ \approx 0 $, which suggests that the system undergoes a rigidity transition without going through any anisotropic state; reminiscent of the approach to the isotropically jammed state~\cite{Bi_nature}. }
   \label{RJ}
   \end{figure}

\begin{figure}[hbtp]
      \centering
   \includegraphics[width=0.5\textwidth]{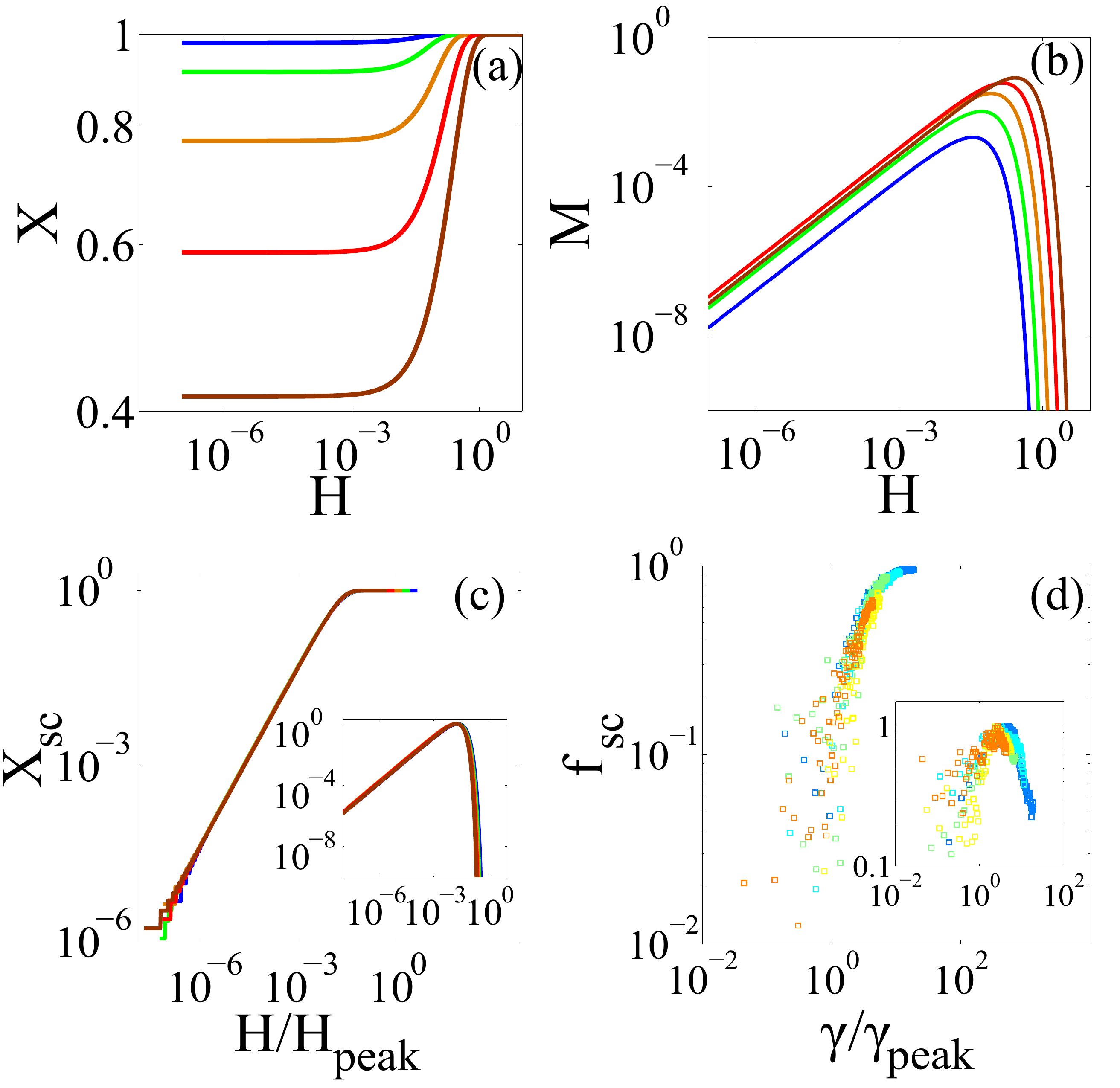} 
   \caption{(Color Online)  \textbf{SJ regime ($ \dn = 0.9$):} (a) $ X (H)$ and (b) $ M (H)$ for  $ 0.2~\mbox{(blue)}  \le R \le 1.5 ~\mbox{(brown)}$ with increments chosen such that $ R/R_J - 1 $ increases logarithmically  between 1 and 10.  (c) Plots  (see text) of $g_X (H/\Hp(R) )$ (main ) and $g_M (H/\Hp(R))$ (inset): $ \Hp \sim (R/R_J - 1)^{1.2} $. (d) Remarkably, we obtain same exact scaling form for $ f_{iso} $ (main) and stress anisotropy (inset) from the SJ experiments~\cite{JieThesis2013}, if we replace $ R/R_J - 1 $ with $ 1 - \phi/\phi_J $ ($ \gamma_{peak}\sim (1-\phi/\phi_J)^{1.2} $).  $(1-\phi/\phi_J) $ varies between 0.02 and 0.09(blue to orange) in the plotted experimental data.     }
   \label{FwdPD}
   \end{figure}

The MF calculation shows that at small $R$ and $\dn  \ge \dc$, $ M(H,R) =0$, and $X(H,R) = 1$ for any $H$.  Physically, this region corresponds to a system in which there are a large number of contacts even at zero driving.  The peak anisotropy vanishes as $M_{peak}(\dn,R) \propto g_{peak} (R/R_J(\dn) -1)$(Fig.~\ref{RJ}),  identifying $ R_J $ as the only characteristic disorder scale in this regime.  The two order parameters, $ M$ and $ X$ are functions of both $H$ and $R$. However, as shown in Fig.~\ref{FwdPD}(c), upon definition of a $R$ dependent characteristic field: $\Hp (R) \propto (R/R_J -1 )^{\delta}$, they obey a scaling form:  $X_{sc}(R,H) = g_X (H/\Hp(R) )$  and $M_{sc} (R,H) = g_M (H/\Hp(R))$, where, $X_{sc}$ and $M_{sc}$ are scaled variables: $ x_{sc}  \equiv \frac{x- x_{min}}{x_{max} - x_{min} } $.   The implication of this result is that in the $\dn > \dc$ regime, the behavior at different disorder strengths $R$ is controlled by the physics of the point ($H=0$, $R_J$), reminiscent of critical phenomena~\cite{goldenfeld_book}.  It was hypothesized by Bi et al~\cite{Bi_nature} that ($ \gamma = 0,\phi = \phi_J $) is a critical point marking the end of a line separating fragile and SJ states. The critical point was characterized by the vanishing of an order parameter, which measures the anisotropy of the stress tensor. The current results, based on the spin model, are consistent with that picture. Numerically, meanfield predicts $\delta = 1.2$, and this exponent collapses experimental data for stress anisotropy and $f_{iso}$ during a forward shear run~\cite{JieThesis2013}, if we identify $R_J$ with $\phi_J$(Fig.~\ref{FwdPD}(d)).

The SJ experiments exhibit the phenomenon of Reynolds pressure~\cite{ren_dijksman}:  pressure increasing quadratically with shear strain at small strains,  with a Reynolds coefficient that depends only on $\phi$ and appears to diverge at $\phi_J$ (Table.~\ref{table:results}). Very general arguments lead to the quadratic dependence of the pressure on shear strain~\cite{tighe_dilat}.  If we make the logical assumption that the pressure increase is determined completely by $f_{iso}$,  and that pressure increases as some monotonic function of $f_{iso}$, and hence $X$, then our model provides a natural explanation for the  observed $\phi$ dependence of pressure. The scaling form of $X(R,H)$ implies that the pressure scales as: $P(R,H)  \sim f(X(R,H)) \propto  g_P (H/\Hp(R))$, where $g_P (x)$ is a scaling function similar to $g_X$ defined above.  The crucial feature of the scaling argument is the vanishing of $\Hp(R)$ as $R \rightarrow R_J$.    From symmetry arguments,  the pressure has to increase as some even function of the shear strain $\gamma$\cite{tighe_dilat}. ($H$ in the model),  $g_P (x)$ increases at least as fast as quadratically with  $x$ for $x << 1$.  Combined with the scaling form, this argument  implies a divergence of the Reynolds coefficient  as some power of  $1/\Hp$, and  therefore, as $ \propto (R/R_J -1 )^{-\delta_P}$, where $\delta_P$ depends on the exponent $\delta$, and the form of $g_P(x)$ for small $x$.    From the perspective of the model, the source of the divergence observed in experiments is, therefore,  directly related to the rapid rise in the number of contacts with shear strain as $\phi$ increases towards $\phi_J$:  a feature that is consistent with experimental observations. 

In the mean-field approximation, there is no hysteresis for $\dn > \dc$.  As we show in Fig.~\ref{Hyst}(a), numerical simulations of the model in 2D exhibit hysteresis in this regime. It is to be noted that, in the simulation, the values of $ \Delta_0 $ and $ R $, which define the SJ regime differ from the MF calculations. However, the overall structure of the phase diagram remains unchanged, as shown in Fig.~\ref{SimPD}.  The model predictions for the scaling of the hysteresis loops (Fig.~\ref{Hyst}(b))  with $R$ are summarized in Table ~\ref{table:results}, and compared to $\phi$ dependence observed in SJ experiments~\cite{ren_dijksman}. 

It is clear from the phase diagram, that the behavior of the model is completely smooth in the regime $\dn > \dc$:  all properties are continuous but sharp changes occur  in the order parameters.  This suggests that the SJ transition in dry grains with frictional coefficient$\simeq 1$~\cite{JieThesis2013} is not a phase transition but a crossover phenomenon at which the contact force network changes continuously both as a function of $\phi$ and $\gamma$.   Preliminary analysis of experiments with lower friction coefficient between grains~\cite{Dong_unpub} suggests that with decreasing friction coefficient $\dn$  approaches $\dc$ from above, which leaves open the possibility of a true transition.

 
\subsection{Scaling and Hysteresis in the DST Regime} {  In the low $\dn$ regime, meanfield analysis predicts multiple solutions to $ M(H,R)$  and hysteresis under cyclic driving.  We can identify three lines based on the multiplicity of solutions:  For $R_m (\dn) \le R \le R_t(\dn)$,  (i) meanfield predicts two solutions for $ M(H=0,R)$ with accompanying hysteresis; (ii)  for $R_m < R < \rd(\dn)$,  multiple solutions appear for $X(H,R)$ leading to multiple hysteresis loops,  as shown in Fig.~\ref{MHyst}. As seen in  Fig. ~\ref{model}(b),  there is a critical point, $ (\dc, R_c)$, marking the end of these three transition lines.    The $R_{DST}$ and the $R_m$ lines are present in numerical simulations in 2D but the $R_t$ line is a meanfield feature.   Simulations exhibit hysteresis over most of the region in Fig. ~\ref{model}(b),  however, their characteristics change at $R_{DST}$, and $R_m$. The $R_m (\dn)$ line marks a discontinuous transition at which the peak anisotropy decreases dramatically, as shown in Figs.~\ref{RJ} and ~\ref{MHyst}.

Identifying $R_m$ with the largest packing fraction, $\phi_m$, at which one can have any flow~\cite{Wyart_Cates}, and $R_{DST}$ with the smallest packing fraction, $\phi_{DST}$,  for the onset of DST, our results imply that two distinct types of force networks are stable in suspensions  with  $\phi_m > \phi > \phi_{DST}$: one with small stress anisotropy and large $f_{iso}$ creating a highly connected network of force-bearing linear structures, and one with larger stress anisotropy and smaller $f_{iso}$.   The networks with large $f_{iso}$ also have large pressures since in our picture, the pressure is determined by $X$.

\begin{figure}[h]
     \centering
  \includegraphics[width=0.5\textwidth]{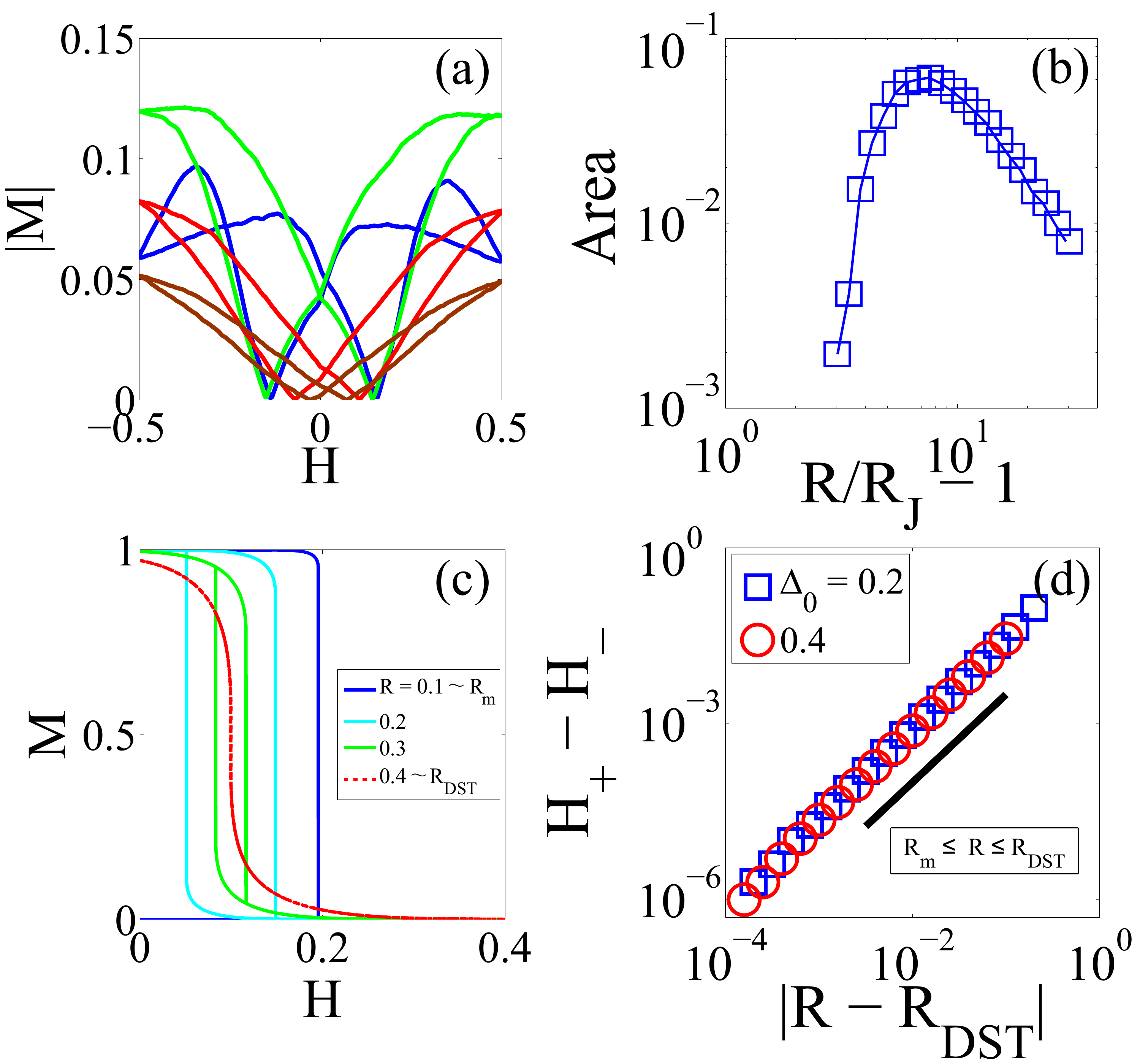} 
  \caption{(Color Online)  \textbf{Hysteresis:} For $ \dn > \dc $, the meanfield solution does not show hysteresis. So, we performed  simulation of the model (Eqn.~\ref{eq:model}) in 2D to compare to SJ experiments, which are in 2D. The details of the simulation method is discussed in the following section. For $ \dn < \dc $,  there is hysteresis even in the meanfield model, and we compare  these results to DST observations in 3D. (a)Hysteresis loops obtained from numerical simulations ($ \dn = 2 $) , for $R/R_J - 1 \in [1,10]$. (b) The area of the hysteresis loops exhibits a non-monotonic behavior and decreases with increasing disorder value beyond a peak. Cyclic shear experiments on dry grains ~\cite{JieThesis2013} show that the size of the hysteresis loops increases with increasing $ \phi $. These experiments do not explore $ \phi $ very near $ \phi_J $, and correspond roughly to $ R/R_J -1 \sim 6-10 $, where the model also predicts increasing size with decreasing $ R $ (increasing $ \phi $). Hence, newer experiments are required to verify this non-monotonic variation of the size of the hysteresis loops.  (c) For $ \dn < \dc $, we observe hysteresis in the meanfield solution for $ R_m \leq R \leq R_{DST} $. The hysteresis loop first appears at $ R_{DST} $ (dotted red line), and increases in size as $ R_m $ is approached from above. Below $ R_m $, no hysteresis loops exist. (d) The size of the hysteresis loop (measured as $ H_+ - H_- $, where $ H_+ (H_-)$ is the maximum (minimum) value of $ H $, where a loop exists) increases as $ R $ is decreased from $ R_{DST} $. The size increases as a power law with exponent $ \frac{3}{2 }$~\cite{Wyart_Cates}.  } 
  \label{Hyst}
  \end{figure}
Meanfield analysis shows that the $X(H,R)$  hysteresis loops span a range $\lbrace H_-  , H_+ \rbrace$, which grows  as $ \left|R - \rd\right|^{\frac 3 2} $ for $R \le \rd $, and $ H_-  \rightarrow 0$  at $ R_m $. These observations are in accord with scaling predictions of hysteresis loops in DST~\cite{Wyart_Cates} if we associate $R_{DST}$ with $\phi_{DST}$.    }

     \begin{figure}[htbp!]
     \centering
     \includegraphics[width=0.5\textwidth]{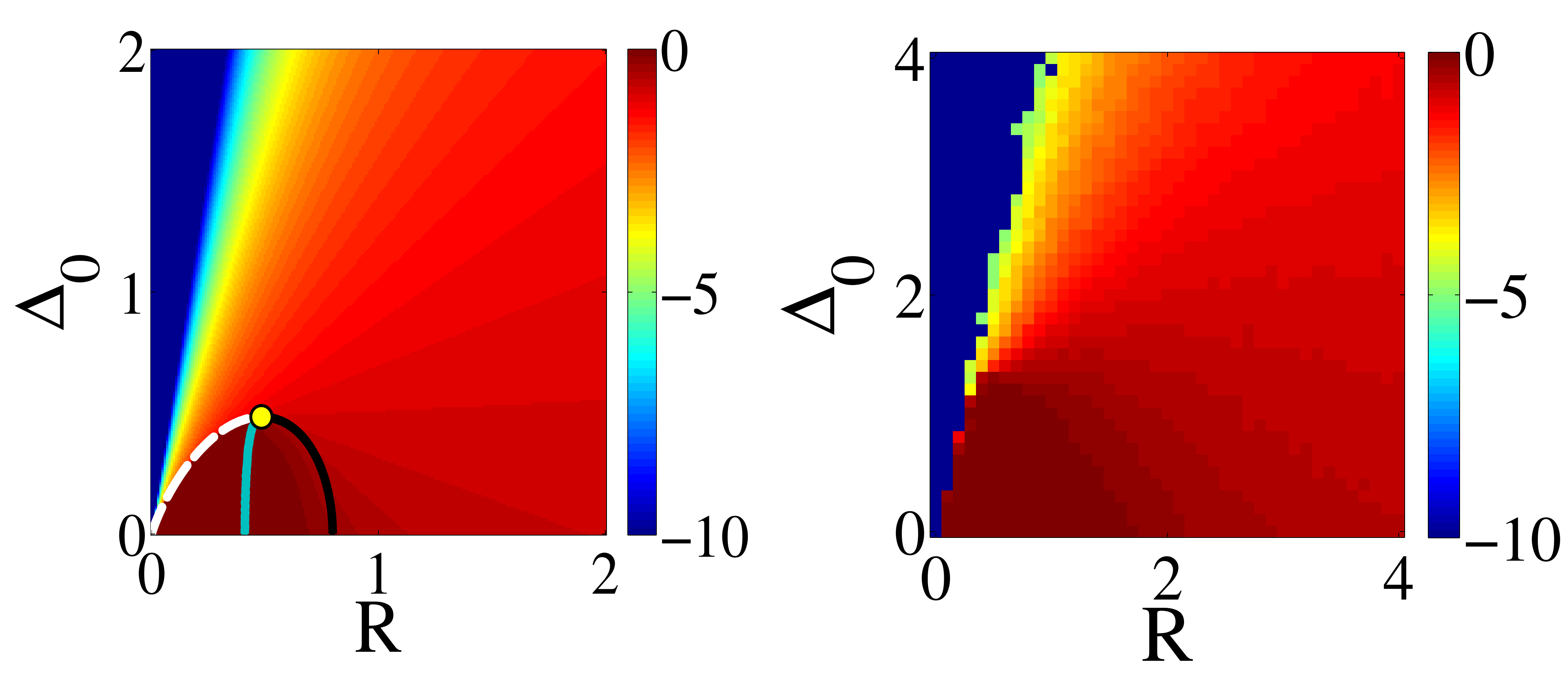}
     \caption{Phase diagram based on the value of $M_{peak}$: MF  (left) compared to the phase diagram obtained from simulations (right).  The colorbar represents the decade in which $ M_{peak} $ lies.  The simulations were performed  on 64$^2$ spins, and averaged over 20 different configurations for each $ \left(R,\Delta_0\right) $.}
     \label{SimPD}
     \end{figure}

\section {Discussion}   We have constructed a driven, disordered, zero-temperature (non-equilibrium) statistical mechanics model, which captures all essential features of  shear-induced rigidity transitions in granular materials and dense athermal suspensions.   
Our analysis highlights the distinct roles played by density and driving in the SJ and DST regimes:  density controls the strength of disorder, whereas the driving field induces rigidity by increasing the concentration of frictional contacts.   
Based on analysis of experiments and simulations, we can assert that the observed phenomenology maps either to the $\dn < \dc$ ($\dn > \dc$)  part (Table.~\ref{table:results}) of the phase diagram based on whether $f_{iso}$  at zero shear is small~\cite{Morris,Mari_long} (large~\cite{JieThesis2013}). Controlling this parameter, for example~\cite{Dong_unpub} by tuning the friction coefficient  of grains,  provides an effective way of controlling where the system lies along the $\dn$ axis in our phase diagram, and probing the behavior near the critical point:$(R_{DST},\dc)$.    

Non-equilibrium critical points of random field models in the Ising class are characterized by avalanche distributions and crackling noise~\cite{Sethna2001}.    Preliminary simulations in 2D indicate that the avalanche distribution exhibits a power law all along $R_{DST}(\dn)$.    Our model is distinguished from the Random Field Ising Model in a crucial way:  $S_i = \pm 1$  can flip back to their original state, mediated by flips to $S_i =0$  even if the  field is being increased monotonically, and the energy at a site does not approach the ``flip'' threshold monotonically.   Recent studies~\cite{Wyart_PNAS} show that this feature affects the yielding transition,  suggesting that our model is  relevant for understanding the yielding of athermal materials.  
We have focused on the shear-induced rigidity aspect of  athermal, particulate systems.   Yielding of the jammed states presumably occurs when the number of frictional contacts is saturated, and shearing does not lead to formation of new  contacts.   We are beginning to explore our model in this regime, where $ X \simeq 1$ and independent of $H$. 

\begin{acknowledgments}
We acknowledge extended discussions with Dapeng Bi, Eric Brown, R. P. Behringer, Jie Ren, Joshua Dijksman, and Dong Wang, and the hospitality of KITP, Santa Barbara, and are  grateful to the Behringer group  for sharing  experimental data.
This work has been supported in part by NSF-DMR 0905880 $\&$ 1409093,  by the W. M.  Keck foundation, and NSF-PHY11-25915
\end{acknowledgments}




\bibliographystyle{apsrev}
\bibliography{reference}

\appendix*

\section {}    
In the following sections we discuss several properties and important aspects of the MF solution. Especially, we discuss the special disorders, which define different boundaries of the MF phase diagram. The special case of $ \alpha = 1 $ trajectories is also discussed here.   

\subsection{Zero disorder behavior}
The energy of a spin $ S_i $ in the zero disorder limit is:
\begin{equation}
E(S_i) = -JMS_i - HS_i + \Delta S_i^2
\end{equation}
If $ S_i = 1 $, and if $ \Delta \geq JM + H $, it will flip to $ S_i = 0 $ state and vice versa. A similar calculation can be done for $ S_i = -1 $. Hence, at zero disorder there is a discontinuous transition from $M=0$ to $M=1$. 
\subsection{Asymptotic behavior of the model}

The asymptotic, large field,  behavior of the model  is governed by the last two terms in Eqn.~\ref{eq:model}. Thus, the effective model governing the behavior at large positive $ H $, with $ \Delta(H) = \alpha \left|H\right|+\Delta_0 $ can be written as: 
\begin{eqnarray}
\lim_{H\to +\infty} \mathcal H &=& -H\sum_i S_i + \Delta(H)\sum_i S_i^2 \label{eqn:Model_Asymp}\\
&=& -H\sum_i S_i + \left(\alpha |H| + \Delta_0\right)\sum_i S_i^2 
\label{eqn:Model_linear}\end{eqnarray}  

The first term on Eqn.~\ref{eqn:Model_Asymp} favors production of $ S = +1 $ when $ H\to +\infty $, whereas the second term favors production of $ S = 0 $ when $ \Delta \to +\infty $. Since $ \Delta $ depends on $ H $, the asymptotic behavior of the model crucially depends on the functional dependence of $ \Delta $ on $ H $, which we refer to as a protocol. For a linear protocol as in Eqn.~\ref{eqn:Model_linear},  which is the only kind we have analyzed, the asymptotic behavior depends on the slope, $ \alpha $. If $ \alpha > 1$, $ \Delta $ dominates $ H $, $ S_i = 0\ \forall i$. Conversely, if $ \alpha < 1 $, $ H $ dominates $ \Delta $, and $ S_i = +1 \ \forall i $. If $ \alpha = 1 $, there is no $ H $ dependence and the asymptotic behavior depends on other terms in Eqn.~\ref{eq:model}. We discuss the $ \alpha \leq 1 $ trajectories in the following section. 

\subsection{Special disorders for $ \alpha > 1 $ trajectories}
The meanfield equations for $ \alpha > 1 $, and $ \Delta_0 < \Delta_c $  admit three lines of transitions which end at a critical point $ \left(R_c, \Delta_c\right) $. These lines are defined by $ R_t  (\dn)$, $ R_{DST} (\dn)$, and $ R_m (\dn)$ in descending order of magnitude (Fig.1(a) in main text). The line $R_t (\dn)$, marks the transition from a single solution for $M(H)$  for $R > R_t (\dn)$ to multiple solutions over a range of $H$ (Fig.~\ref{MHyst}), whereas the line $ R_m(\dn) $ marks the transition from multiple solutions for $ M(H)$ with $M_{peak} \simeq 1$  for $R = R_m^+ (\dn) $  to  a single solution with $M_{peak} \simeq 0$ for $R =R_m^- (\dn) $, as shown in Fig.~\ref{MHyst}. Notably,  $M_{peak}$ has a discontinuity at $R_m(\dn)$ with the discontinuity increasing as $\dn \rightarrow 0$, as seen in Fig.~\ref{RJ}. The transition at $ R_t (\dn)$ is continuous. The transition lines, $ R_t (\dn)$ and $ R_m (\dn)$, can be calculated analytically from the mean field equations and yields: $ R_t (\dn) = \sqrt{\frac{-\Delta_0^2}{W\left(0,-\frac{\pi\Delta_0^2}{2}\right)}}  $, and $  R_m (\dn)= \sqrt{\frac{-\Delta_0^2}{W\left(-1,-\frac{\pi\Delta_0^2}{2}\right)}}  $. Here $ W(k,x) $ is the product log function, also known as Lambert's $ W $ function. 

The transition at $R_{DST} (\dn)$ is a unique feature of our model and marks the onset of multiple solutions to $X(H)$, accompanied by system-size avalanches  in which spins flip from $ \pm 1$ to $0$. $ R_{DST} (\dn)$ is difficult to calculate analytically, and the line shown in Fig. 1(a) of the main text has been obtained numerically. Apart from $ \Delta_0 $ very close to $ \Delta_c $, $ R_{DST} (\dn)\approx 0.4$. Fig. ~\ref{MHyst} illustrates the behavior of the system near these special disorders by comparing the $ M $-hysteresis.     In the main text, these special disorders have been related to special packing fractions  relevant to the DST transition.

\begin{figure}[htbp!]
\centering
\begin{center}
\centerline{\includegraphics[width=0.5\textwidth]{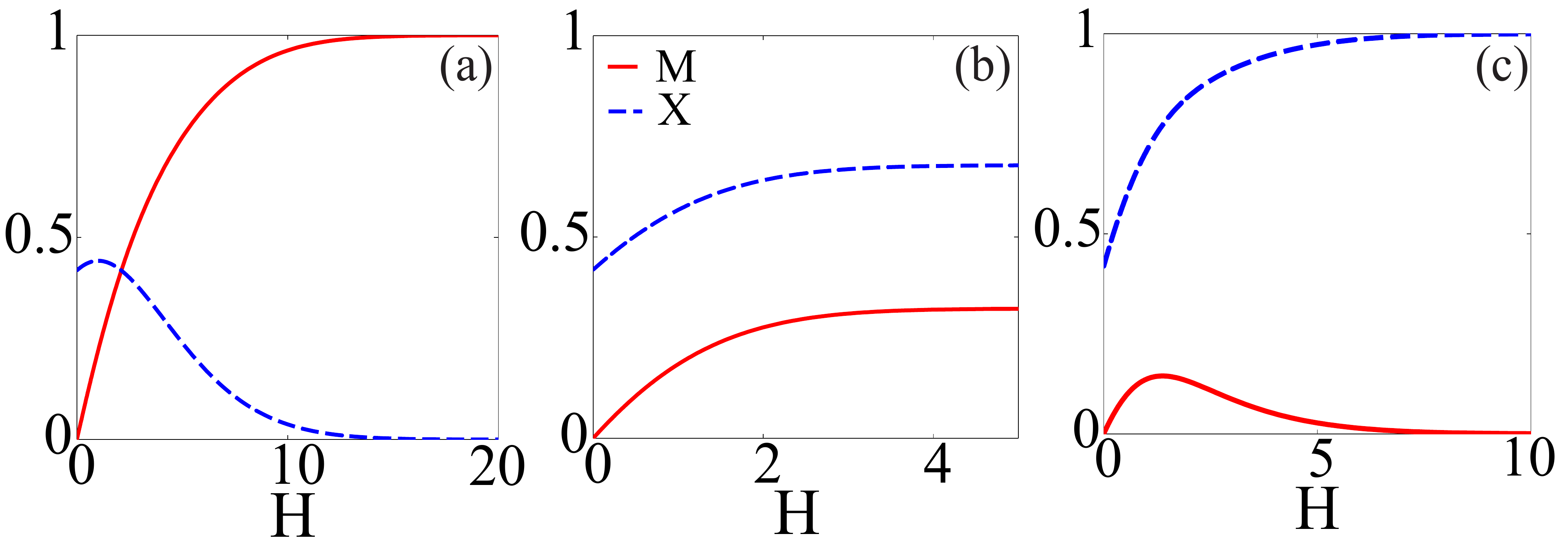}}
\caption{Comparison of trajectories with different $ \alpha $, $ \Delta_0 > \Delta_c $. The asymptotic ($ H>>H_{peak} $) dynamics is governed by $ \alpha $. $ M $ monotonically increases to 1 while $ X $ decreases to zero for $ \alpha < 1 $ trajectories (a). The exact  opposite trend is observed for $ \alpha > 1 $ trajectories (b). For $ \alpha = 1 $, both $ M $ and $ X $ increases monotonically and saturate to a value less than 1 (c). The saturation value depends on disorder.}
\label{alpha}
\end{center}
\end{figure}

\begin{figure}[htbp!]
\centering
\includegraphics[width=0.5\textwidth]{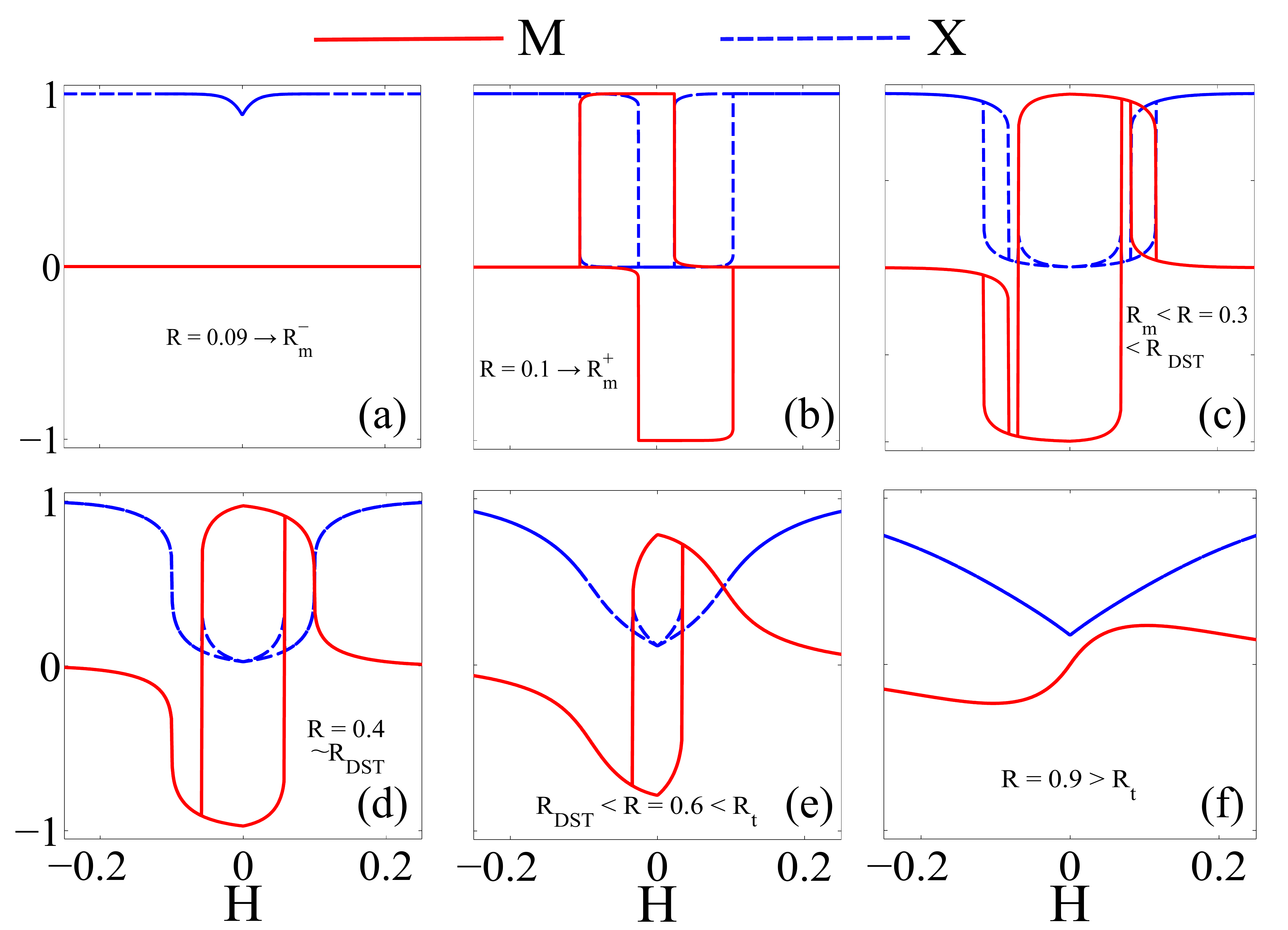}
\caption{ Hysteresis of $ M $ and $ X $ illustrate behavior of the model for different disorders ($ \alpha = 4, ~\mbox{and} ~ \dn = 0.2 $). $ R_m \sim 0.097 $, $ R_{DST} \sim 0.4 $, and $ R_c \sim 0.8 $ are the special disorders for $ \dn = 0.2 $ (see specialk disorders subsection). At $ R_m $, the peak value of $ M$, $ M_{peak} $, changes discontinuously (Fig. ~\ref{RJ}(a)). (a)  $R = 0.09 $ (just below $ R_m $), (b) $ R = 0.1 $ (just above $ R_m $), (c) $ R = 0.3 $ (below $R_{DST}$), (d) $ R = 0.4 $ ($ R_{DST} $), (e) $ R = 0.6 $ ($ R_{DST} < R < R_t $), and (f) $ R = 0.9 $ ($ R > R_t $).      }
\label{MHyst}
\end{figure} 

\subsection{$ \alpha \le1 $, $\Delta > \Delta_c$ trajectories}

For $ \alpha < 1 $, $M$ increases and  $X$  decreases as $H$ is increased, indicating  that $S_i = \pm 1$ proliferate (Fig.~\ref{alpha} (a)).  This trajectory is, therefore,  not relevant to shear induced rigidity where grains with three or more contacts ($S_i = 0$) proliferate, as the system is driven towards jamming.

\begin{figure}[htbp!]
\centering
\includegraphics[width=0.5\textwidth]{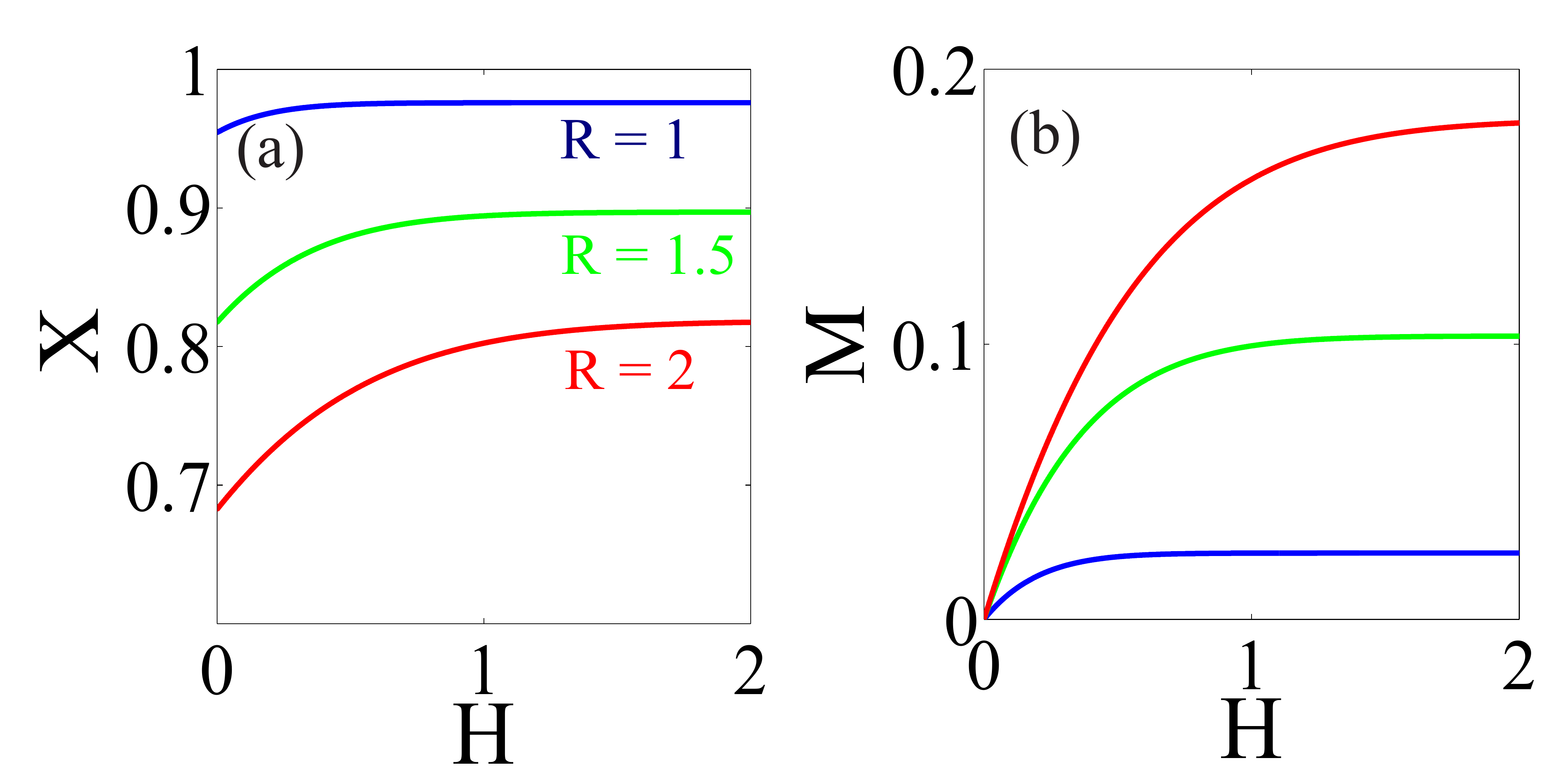}
\caption{$ X $ (a) and $ M $ (b) as a function of field $ H $ for $ \alpha = 1 $ ($ \Delta_0 = 0.9 $) trajectories for a few typical disorder strength; obtained from meanfield. Both order parameters increase monotonically, and saturate to a value less than 1. The saturation value depends on $ R $. For $ M $, the saturation value increases with $ R $ while for $ X $ it decreases. }
\label{alpha1MX}
\end{figure}

Lying between $ \alpha < 1 $ and $ \alpha > 1 $ trajectories, $ \alpha = 1 $ trajectories exhibit an interesting dynamics (Fig.~\ref{alpha}(b)). Since the chemical potential, $ \Delta(H) $ changes at the same rate as $ H $, the applied field,  the production of $ \pm 1 $ spins favored by  $ H $ competes equally with the production of $ 0 $ spins favored by $ \Delta $.    For $\alpha > 1$ trajectories,  the magnetization $M(H)$ starts to decrease with increasing $H$ for $H > H_{peak} (R)$, as depicted in Fig. 2 of the main text.  In contrast, for $\alpha=1$,  we observe that both the magnetization $ M $, and the fraction of zero spins $ X $ asymptote to a disorder-dependent values $M_{sat}$ and $X_{sat}$  for $ H >> H_{peak} $.  As $ R $ increases, $ M_{sat} $ increases while $ X_{sat} $ decreases as shown in Fig. ~\ref{alpha1MX}.

\begin{figure*}[ht!]
\centering
\includegraphics[scale =0.2]{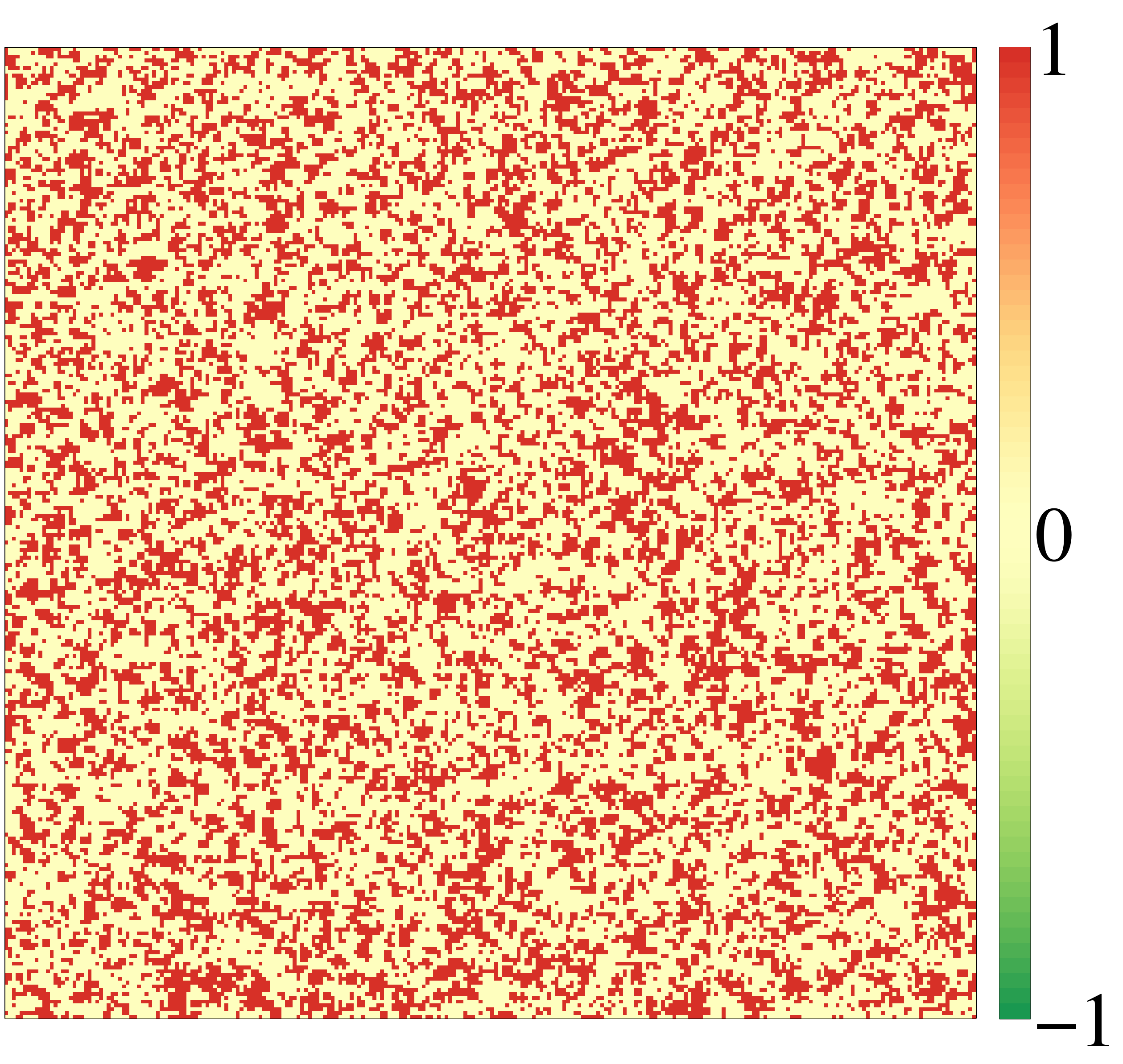}
\caption{Numerically obtained asymptotic $ (H >> H_{peak}) $ spin configuration for $ \alpha = 1 $ trajectories. $ \Delta_0 = 2 > \Delta_c $ and $ R = 2 $.  } 
 
\label{alpha1}
\end{figure*}

Simulations of the model (Eq. ~\ref{eq:model}) in 2D, using zero temperature Monte Carlo dynamics,  show that the asymptotic  states for $\alpha=1$ have a non-trivial spatial distribution of spins. As shown in Fig.~\ref{alpha1}, there is micro-phase separation between $ \pm 1 $ and $ 0 $ spins. This spatial structure is reminiscent of  shear bands observed in shear jamming experiments [3].

\end{document}